\documentclass{emulateapj} 
\shorttitle{Circumbinary disk around CoKu Tau/4}
\shortauthors{Ireland and Kraus}
\begin{document}

\title{The Disk Around CoKu Tau/4: Circumbinary, not Transitional
\altaffilmark{1} }

\author{Ireland, M.J.}
\affil{Division of Geological and Planetary Sciences, California Institute of
  Technology, Pasadena, CA 91125}
\email{mireland@gps.caltech.edu}
\and
\author{Kraus, A.L.}
\affil{Division of Physics, Mathematics and Astronomy, California Institute of
  Technology, Pasadena, CA 91125}

\altaffiltext{1}{Data presented herein were obtained at the
W.M. Keck Observatory, which is operated as a scientific partnership
among the California Institute of Technology, the University of
California and the National Aeronautics and Space Administration. The
Observatory was made possible by the generous financial support of the
W.M. Keck Foundation. }

\begin{abstract}
CoKu Tau/4 has been labeled as one of the very few known transition
disk objects: disks around young stars that have their inner disks cleared of
dust, arguably due to planetary formation. We report aperture-masking
interferometry and adaptive optics 
imaging observations showing that CoKu Tau/4 is in fact a near-equal
binary star of projected separation $\sim$53\,mas ($\sim$8\,AU). The spectral 
energy distribution of the disk is then naturally explained by inner
truncation of the disk through gravitational interactions with the
binary star system. We discuss the possibility that such ``unseen''
binary companions could cause other circumbinary disks to be labeled as
transitional.
\end{abstract}

\section{Introduction}

The lifetime of disks around young stars sets the time-scale for 
giant planet formation. Most young stars have disks that are optically
thick at all wavelengths or have no observable disk at all, while a
small subset of young stars are in an intermediate state, thought to
be in ``transition'' \citep{Skrutskie90}. Many of these intermediate
objects have optically thick ``cold disks'' as seen by large excesses
at $\sim$30\,$\mu$m, but much smaller or non-existent near- and mid-
infrared excesses \citep{Brown07}.

%For the vast
%majority of objects, the presence of an inner disk, seen by a
%near-infrared excess, correlates with the the spectroscopic
%classification of a classical T~Tauri star, while the lack of an
%optically-thick inner disk means an object is a weak-lined T~Tauri star
%\citep[e.g.][]{Hartmann05}. 

One model of this transition process is
one where the inner disk is cleared of material, leaving an
inner hole and low accretion rates, followed $\sim10^5$ years later by
a clearing of the outer disk. This inner hole could be caused by natural
processes of grain growth and/or photo-evaporation
\citep{Alexander06}, or truncation of the disk due to dynamical 
interactions with companions \citep{Lin79}.
Although the most exciting interpretation for the development of gaps
is arguably truncation due to planetary formation, it is also possible 
to dynamically truncate disks due to stellar companions 
\citep[e.g][]{Jensen97,Beust05,White05}. An example of
this is CS~Cha, which was announced as a transitional disk
\citep{Espaillat07} shortly after its discovery as a circumbinary disk
\citep{Guenther07}. Some authors \cite[e.g.][]{Furlan07} have
considered the term ``transitional'' to be an observational term that
means the spectral signature of a cleared inner disk
rather than an evolutionary term that implicitly applies to single stars. 
We use the term ``cold disk'' for this purpose.

CoKu Tau/4 is a weak-lined M1.5 T~Tauri star \citep[H-$\alpha$ equivalent
  width 1.8\,$\AA$, ][]{Cohen79} in the Taurus star-forming region
(1$\sim$2\,Myr, $\sim$145\,pc).
CoKu Tau/4 was discovered to have a large $\sim$20-30\,$\mu$m excess
and no excess at wavelengths $<8$\,$\mu$m through {\em Spitzer Space
  Telescope} Infrared Spectrograph observations \citep{Forrest04}. The
disk has been modeled as having an inner hole of radius $\sim$10\,AU
\citep{DAlessio05}, with 
suggestions that this hole is due to a giant planet \citep{Quillen04}.

In this paper, we describe near-infrared aperture-masking
interferometry and imaging observations that demonstrate that CoKu
Tau/4 is a near-equal mass binary star system. We show that the
predicted inner hole size from dynamical models is comparable to, but larger
than, that predicted from radiative transfer models. 
Finally, we discuss whether other so-called ``transitional'' disks could 
be circumbinary disks. We conclude that for candidates in Taurus, much of the 
mass ratio-separation space for stellar companions can by ruled out by 
existing observations, but that definitively ruling out the possibility
of binarity is generally difficult for individual so-called ``transitional'' 
disks. 

\section{Observations}
\label{sectObservations}

CoKu Tau/4 was observed with the NIRC2 camera behind Adaptive Optics
(AO) at the Keck II telescope on 2007 Nov 23 as
part of an ongoing aperture-masking survey of nearby young
star-forming associations. Aperture-masking
interferometry \citep[e.g.][]{Tuthill00} is a well established
technique for achieving the full diffraction limit of a single
telescope, recently applied to observations behind adaptive optics
systems \citep[e.g.][]{Lloyd06,Kraus08a}. A 9-hole mask was placed in a
filter wheel near a pupil plane in the NIRC2 camera, enabling
interference fringes on 36 baselines to be simultaneously recorded on
the camera's imaging array. The observations of CoKu Tau/4 consisted
of two image sets taken through a K' filter,
each with eight 20 second exposures, calibrated by 
two interleaved image sets of CX~Tau. The airmass of CoKu Tau/4
observations varied from 1.28 to 1.38, while the airmass of CX Tau (6
degrees from CoKu Tau/4) varied from 1.32 to 1.45 over the 45 minutes
it took to complete the observations. From the Fourier transforms of the target
and calibrator's images, squared-visibilities and closure-phases were
extracted. The squared visibilities of CoKu Tau/4 were calibrated by
dividing by the squared visibilities of the calibrator, and the
closure-phases were calibrated by subtracting the closure-phases of
the calibrator (which are non-zero due to instrumental effects). These
observations were made 
in poor seeing conditions (uncorrected seeing 1-2\arcsec in
K-band), so our observations only achieved typical sensitivities of
$\sim$50:1 at the diffraction limit, less than the $\sim$200:1 we
achieved in Upper Sco \citep{Kraus08a}. These high sensitivities are
primarily due to the non-redundant nature of the aperture-mask, which
means that calibration is independent of wavefront structure on scales
larger than a single sub-aperture ($\sim$1\,m).

\begin{figure}
 \plotone{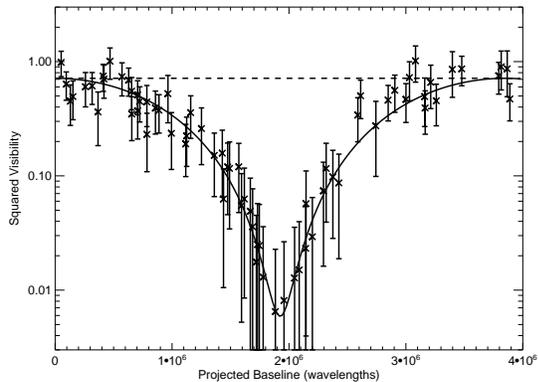}
 \caption{Calibrated squared-visibilities for CoKu~Tau~4, plotted
   against baseline projected along a position angle of 306 degrees. 
   A binary model with contrast 1.25 and separation 54\,mas is
   over-plotted. The dashed line shows the visibility expected for a
   single star.}
 \label{figV2}
\end{figure}

\begin{figure}
 \plotone{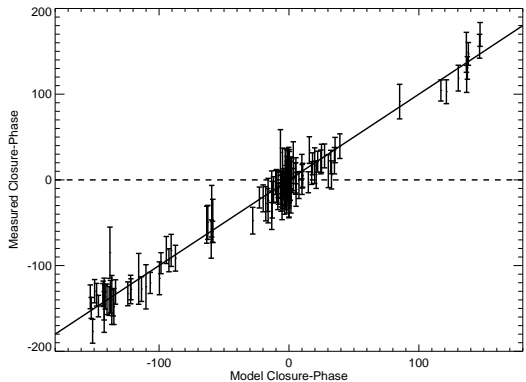}
 \caption{Measured closure phases for CoKu~Tau~4 as a function of
   modeled closure phases, assuming that CoKu~Tau~4 is a 1.25:1 binary
   with separation 54\,mas at a position angle of 306 degrees. The
   dashed line shows the zero closure-phase expected for a single star.}
 \label{figCP}
\end{figure}

\begin{figure}
 \plotone{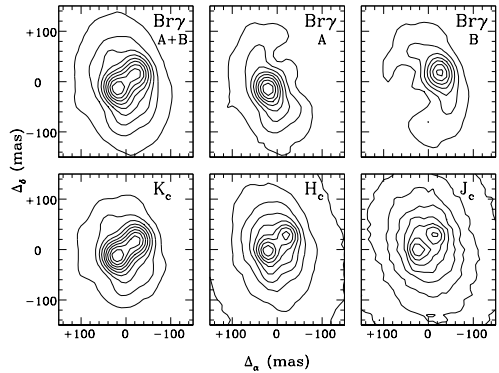}
 \caption{Top line: Contour plots of the original AO image, only
   the A component (best-fit B solution subtracted), and only the B component
   (best-fit A solution subtracted) in the Br$\gamma$ filter,
   demonstrating the PSF fitting technique. Bottom line: AO images in
   the K$_{cont}$, H$_{cont}$ and J$_{cont}$ filters, clearly showing
   the binary despite the low Strehl. The contours are drawn at 10\%
   to 90\% in intervals of 10\%. } 
 \label{figIm}
\end{figure}

\begin{deluxetable}{lcrrr}
\tabletypesize{\scriptsize}
\tablewidth{0pt}
\tablecaption{Direct Imaging Observations}
\tablehead{\colhead{Filter} & \colhead{Wavelength} &
\colhead{$\rho$} & \colhead{PA} & \colhead{$\Delta$}
\\
\colhead{} & \colhead{Range ($\mu$m)} &
\colhead{(mas)} & \colhead{(deg)} & \colhead{(mag)}}
\startdata
\multicolumn{5}{l}{Aperture Masking}\\ %320s integration
$K'$     &1.948-2.299 & $54.1\pm0.3$ & $306.7\pm 0.4$ & $0.23\pm0.01$ \\
\multicolumn{5}{l}{PSF Fitting} \\
$H_{cont}$&1.569-1.592 & $53.6\pm0.7$ & $306.3\pm0.3$ & $0.24\pm0.04$ \\
$Br\gamma$&2.152-2.185 & $53.3\pm0.2$ & $307.7\pm0.7$ &$0.25\pm0.01$ \\
$K_{cont}$&2.256-2.285 & $53.6\pm0.2$ & $307.1\pm0.2$ &$0.21\pm0.01$ \\
\multicolumn{5}{l}{Bispectrum Analysis} \\
$J_{cont}$ &1.203-1.223 & $51.7\pm0.9$  & $305.8\pm0.6$ & $0.23\pm0.06$  \\
$H_{cont}$ &1.569-1.592 & $53.3\pm0.2$  & $306.9\pm0.2$ & $0.21\pm0.02$ \\
$Br\gamma$ &2.152-2.185 & $53.9\pm0.2$  & $306.4\pm0.2$  & $0.19\pm0.01$ \\
$K_{cont}$ &2.256-2.285 & $53.6\pm0.2$  & $306.0\pm0.2$ & $0.20\pm0.01$ \\
\enddata
\label{tabImaging}
\end{deluxetable}

We found that only a binary model provided
a good fit to the aperture-masking data, with a near-equal binary well
within our detection limits. The calibrated squared visibilities and
closure phases are shown in 
Figures~\ref{figV2} and \ref{figCP}, demonstrating that a binary
solution provides an excellent fit to the data. The binary solution
based on the aperture-masking data is given in Table~\ref{tabImaging}. 
%a $54.1\pm0.3$\,mas separation
%at a position angle of $306.7\pm 0.4$ and a contrast of
%$0.23\pm0.01$\,magnitudes.  

Preliminary analysis of the masking data indicated that the binary
companion's separation and flux ratio would allow it to be resolved
with direct imaging, so we obtained regular AO images
on the following night under slightly better observing conditions
(uncorrected seeing $\sim$1\arcsec in K-band) at an airmass of 1.06.   
These observations were obtained with NIRC2 in the narrow
camera mode ($\sim$10 mas pix$^{-1}$), and since the target is very
bright in the near-infrared, we used the narrow-band filters
$J_{cont}$, $H_{cont}$, $K_{cont}$, and $Br\gamma$.  Total exposure
times were 40\,s in all filters except for $J_{cont}$, where twice as
many images were taken due to the very low Strehl, for a total
exposure time of 80\,s. We show the images in
Figure~\ref{figIm}. Despite the low Strehl ratio and the elongation of
the speckle halo due to a dominant wind direction, the binary is
clearly resolved in all images.

We extracted photometry and astrometry from the direct AO imaging 
of these sources using the
PSF reconstruction technique described in \citet{Kraus07b},
which iteratively fits a template PSF to the primary and then
subtracts the secondary to fit an improved estimate of the
primary. This routine was implemented using the ALLSTAR routine in the
IRAF package DAOPHOT (Stetson 1987), and based on previous experience
with NIRC2 imaging, we chose to model the PSF using a Gaussian core
with Lorentzian wings. This method is illustrated in
Figure~\ref{figIm}. In addition to this method, we used a recursive-phase
algorithm \citep[e.g][]{Lohmann83} to reconstruct Fourier Phase from  
bispectrum phase, and then fitted a binary model to the reconstructed phase of
the image. 
In Table~\ref{tabImaging}, we summarize the relative astrometry
and photometry that we measured from our direct imaging observations. 
The consistency in separation and position angle at the 1\% level in H and 
K bands and the 5\% level in J band despite very different PSFs and diffraction
effects demonstrates the reliability of our results. However, the
small reported statistical errors in Table~\ref{tabImaging} are not
entirely consistent internally, requiring additional
systematic errors to be added. Therefore, we assign a weighted mean
solution of separation $53.6\pm0.5$\,mas and a position angle of
$306.4\pm0.6$\,degrees. The contrast ratio in J, H and K  is
consistent with there being no significant difference in reddening
between the components.

\section{System Properties}

\subsection{Stellar and Binary Properties}
\label{sectProperties}

Both components have near infrared colors that are consistent with other young
Taurus members of spectral type $\sim$M1-M2, suggesting that they have
similar temperatures and that neither has a significant $K$ band
excess due to a hot inner disk. In order to model the stars and the
disk, it is important to know the bolometric 
luminosity of the system, and hence the reddening.
To estimate the reddening, we used the following data: intrinsic colors
reported in \citet{Bessell88} and \citet{Bessell90} interpolating
between giants (log$(g) \sim 1$) and dwarfs (log$(g) \sim 5$) at an
assumed log$(g)$ of 3.7, JHK photometry from 2MASS,  V-band photometry
averaged between \citet{Cohen79} and \citet{Hanson04} with an assumed
0.1\,mag overall uncertainty, and an M1.5 spectral classification 
from \citet{Cohen79} with an assumed 1 sub-class uncertainty. We arrive at
$E_{V-K}=2.3\pm0.3$ and $E_{J-K}=0.58\pm0.04$.  The ratio $E_{V-K}/E_{J-K} =
4.0\pm0.6$ is significantly lower than the 5.6 typical of Taurus
\citep{Whittet01}, suggesting a reddening law more typical of a dense
core. Based on $E_{V-K}$, the reddening $A_V$ is $2.5\pm0.3$ using the
relationship $A_V \approx 1.1 E_{V-K}$ from \citet{Whittet01}, but given
the uncertainties in the reddening law, $A_V=3$ as used by
\citet{DAlessio05} is certainly consistent. 
%NB: A small K-band excess doesn't make sense, because even with
%optically-thin dust, a ~0.1 mag K-band excess gives ~1 mag at the
%short end of IRS

%NB2: Kirkpatrick (1991) has a good spectral atlas of M stars.

%Using Lynne's Reddening table: We arrive at a
%reddening of $A_V=2.59\pm0.12$ from $V-K$, and  $A_V=3.56\pm0.22$ from
%J, H and K photometry alone. An uncertainty of 1 sub-class in spectral
%type corresponds to an additional reddening uncertainty of
%$\sim$0.25 for $V-K$ based reddening, with a negligible difference to
%the reddening based on $J$, $H$ and $K$ magnitudes. We do not have a
%way to reconcile these numbers, but note that \citet{Flaherty07} find
%that in dense clouds typical of star forming regions, near infrared
%reddening laws differ substantially from those based on interstellar
%dust.

%NB, if we were to B-band photometry from \citet{Hanson04}, then that
%is consistent with an _unreddened_ star!

Absolute stellar properties like
masses can be difficult to estimate because pre-main-sequence
evolutionary models are not well-calibrated, but based on the
prescription described in \citet{Kraus07a}, both
components are likely to have masses of $\sim$0.5-0.6
$M_{\sun}$.  Using a 145\,pc distance to Taurus, the system has an
absolute K magnitude of 2.55, which implies absolute K magnitudes of
3.2 and 3.4 for the components. According to the models of
\citet{Baraffe98}, assuming an effective temperature of $\sim$3600\,K
applicable to M1.5 stars \citep{Luhman99}, the stars are 4\,MYr old
and are of mass $\sim$0.52 and $\sim$0.62\,$M_\sun$. If the effective
temperature is closer to 3500\,K, the stars would be of slightly lower
mass and $<3$\,Myr old, consistent with the canonical age of Taurus. 
The observational scatter in the Taurus HR diagram
\citep[e.g][]{Luhman04} is consistent with CoKu Tau/4 being either
single as assumed by previous authors or an equal brightness double.
The mass ratio, however, should be more robust than the masses or age
(as it is only weakly dependent on reddening), so we can assert with
greater confidence that $M_B/M_A=$0.85$\pm$0.05.

The minimum possible semi-major axis ($a$) for the orbit is
the apparent separation divided by (1+$e$), which is 7.8/(1+$e$)\,AU
based on an assumed distance 
of 145\,pc for Taurus, giving an orbital period $>$10 years. 
Based on considerations of dynamical disk
truncation in Section~\ref{sectModel}, it is likely that the system
was observed near maximum elongation.

\subsection{The Circumbinary Disk}
\label{sectModel}

%NB Roche Lobe really has nothing to do with this, at only 0.764a. 
%2/(2r-a) + 2/(2r+a) - r^2/a^3 = 4/a
%If instead, emission were spread over a range
%of radii (e.g. in a flared disk with a curved wall), one would expect
%this to drive the inner-most radius of emission inwards. Furthermore,
%\citet{DAlessio05} find an insignificant contribution from dust inside
%the inner wall to the mid-infrared spectrum. We should therefore
%consider their radii to be model-dependent outer limits of the disk
%truncation radius. 

We expect the inner radius of the CoKu Tau/4 disk to be set by tidal
truncation. To simplify our discussion, we will assume that the
disk is roughly co-planar with the binary and (as described in
Section~\ref{sectProperties}) that the binary components have similar
masses. In this geometry, \citet{Artymowicz94} calculate that the 3:1
orbital resonance at $R_D\sim2.08a$ is opened for all but improbably
small eccentricities ($e\ga0.03$). \citet{Beust05} predict a
gap size of $>2.6a$ for eccentricities $>0.1$ for GG~Tau, which has a
similar mass ratio to CoKu Tau/4 (and find this not to be large enough
for that system). For very large eccentricities ($e\ga0.6$), larger
resonances open and predicted gap sizes from \citet{Artymowicz94} are
in excess of $R_D\sim3a$ but are viscosity-dependent. The closer
near-equal spectroscopic binary HD~98800B \citep{Akeson07} is an
example of a system with high eccentricity that should have opened
large resonances in the disk: but the radiative transfer modeling has
too many uncertainties to clearly say which resonance has opened. 

%NB St 34 is near equal but old, and is Hartmann05b

%The term of order $e^1$ for
%the 4:1 resonance vanishes for equal-mass binaries
%\citep{Artymowicz94}, so we will not consider this here. Note,
%however, that the opening of this and higer order resonances will be
%dependent on disk viscosity considerations, and for the binary GG~Tau,
%which has a similar mass-ratio to CoKu Tau/4, \citet{Beust05} predict
%a gap size of $>2.6a$ for eccentricities $>0.1$. 
%See equation 24 in Artymowicz. There is a (1-2\mu) term.

Based on the minimum semi-major axis of 7.8/(1+$e$)\,AU from
Section~\ref{sectObservations}, we arrive at minimum disk
truncation radius of $\sim$16\,AU if CoKu Tau/4 is a low ($e\sim0.1$)
eccentricity binary, or $\sim$13\,AU if CoKu Tau/4 is a high ($e\ga0.6$)
eccentricity binary. These numbers are
discrepant from the 9-12\,AU truncation radius predicted by the
detailed spectral modeling of \citet{DAlessio05}, who examined
radiative transfer models consisting of a geometrically thin wall and
a range of dust types. We will examine several ways to resolve this
discrepancy in turn: the differing radiation pattern of a binary
versus a single star, errors in the stellar luminosity, and the
possibility of a substantially different radiative transfer model.

%Note that this discrepancy has the opposite sign to that
%for the well-known circumbinary disk around 
%GG~Tau \citep{Beust05}, where the disk appears too large given the
%binary separation. 

A binary star has a different radiation pattern to a single star. In
particular, on a circular ring of radius $R_d$ centered on the binary center
of mass, flux is greatest at points on this ring aligned with the
axis of the binary. For an 8\,AU binary (each component 4\,AU from the
center of mass) and a 14\,AU ring, this geometry means that the flux
from the binary is between 0.92 and 1.28 times the flux from a single
star with the same total luminosity. We will assume that the
$\sim$145\,K modeled by \citet{DAlessio05} represents the maximum
temperature seen in the ring and this temperature determines
$L_S/R_D^2$. Here $L_S$ is the total stellar luminosity and $R_D$ is
the dust radius from the center of mass. Therefore, a 28\% increase in
flux $F$ could represent a $\sim$13\% increase in the ring radius from
radiative transfer modeling. The temperature contrast around the ring
for optically thin dust with opacity $\kappa_\lambda \propto
\lambda^{-1}$goes as $F^{1/5}$, meaning that the temperature around
the ring ranges from $\sim$136\,K to 145\,K. Gray dust would vary from
$\sim$133\,K to 145\,K. Although this contrast is only moderate, it
may present an additional challenge in detailed spectral fitting, as
the models of \citet{DAlessio05} required emission from a single
temperature only. 

%so the apparently excellent fit of the
%CoKu~Tau/4 spectrum with a dust at a single radius in \citet{DAlessio05} is
%not as compelling evidence for a single optically-thin dust
%temperature as it initially might seem.

% 0.8 power of luminosity. 1.28 versus 0.92.
%So a factor of 1.3 between coolest and warmest parts to the disk.

%A 14\,AU radius ring that is co-planar with this
%same binary in turn receives a 10\% increase in flux over a
%single star of equivalent luminosity. This effect would cause the disk
%radii of \citet{DAlessio05} to be underestimated by $\sim$5\%.

One way to reconcile the gap radius from radiative transfer modeling
with that from
dynamical considerations would be if the luminosity of CoKu Tau/4 is
significantly underestimated. If we keep the dust temperature and
therefore $L_S/R_D^2$ constant, increasing $L_S$ increases the model
$R_D$. An uncertainty in distance due to
the depth of Taurus \cite[e.g.][]{Torres07} can not reconcile this
discrepancy, as both $R_D$ determined from radiative transfer
modeling and the minimum $R_D$ determined by dynamical
considerations both increase linearly with the distance estimate.
However, an uncertainty in the reddening of CoKu Tau/4 translates to a
luminosity uncertainty. The probable reddening calculated in
Section~\ref{sectProperties} is lower than that used by
\citet{DAlessio05}, but uncertainties are considerable, and it may be
possible that a larger bolometric reddening could increase the gap
radius derived from radiative transfer modeling.

%2/(14^2 + 4^2) versus 1/10^2 + 1/18^2

%!!! I did say 60% what is the real number...? The infrared reddening
% would produce a 1 mag V-band excess, and a 2.4 mag B-band excess
%... which is significantly different from the value in
%\citet{Cohen79} of $0.94\pm0.29$.

In order to independently estimate the dust radius from radiative
transfer modeling and appreciate possible further complexities, we
will examine a very simple model. This 
model contains small grains of amorphous olivine (i.e. the limit where
the grain radius is much smaller than the blackbody emission peak),
with optical constants from \cite{Dorschner95}. We assign a dust temperature
$T_D$=145\,K and a stellar luminosity of 0.61\,$L_\sun$ as in 
\citet{DAlessio05}, and a stellar temperature of an M1.5 star of
$T_S=$3600\,K \citep{Luhman99}. The stellar radius for each of the two
stars is 3.7\,$R_\sun$, equivalent to a single star of radius
5.2\,$R_\sun$. We assume a blackbody spectrum for
the star\footnote{Assuming instead a NextGen model spectrum from
  \citet{Hauschildt99} at 3600\,K and log($g$)=4.0 reduces $R_D$ by
  2\,AU, but has far too much TiO absorption when compared to an M1.5
  spectrum}. The dust is assumed optically thin (which could represent
the outer layers of an optically-thick disk), so that the dust radius
is then given by:
%3600K from Adam's Coma Ber paper.

%This is also in calc_redenning.script
\begin{equation}
  R_D = R_S \sqrt{\frac{\int \kappa_\lambda B_\lambda(T_S) d\lambda}
    {2 \int \kappa_\lambda B_\lambda(T_D) d\lambda} }
\end{equation}

In this simple model, the dust radius is $R_D=$15\,AU, roughly consistent with
disk truncation radii from dynamical considerations. With the many
possible parameters to tweak in a more complete model, and the
possibility of the optically-thin dust being additionally heated by
surrounding dust, increasing $T_D$ and driving the
model radius outwards, it appears likely that a detailed radiative
transfer model could indeed be made to match the dynamical truncation
radius. The edge in the spectrum of disk
emission from CoKu Tau/4 at $\sim$9\,$\mu$m coincides with a sharp
rise in silicate emissivity, meaning that the interplay between
temperature, extinction, geometry and dust composition is all the more
complex for this source. We hope that this preliminary discussion will
motivate other authors to examine possible models in more detail.

\section{Discussion and Conclusions}
\label{sectConclusion}

The circumbinary nature of the CoKu Tau/4 disk begs the question: are
other so-called transition disks also likely circumbinary in nature?
The other well-known cold disks in Taurus, 
DM~Tau, LkCa~15, UX~Tau and GM~Aur, are also generally assumed to be
single stars.  Although all these systems differ from one another,
CoKu Tau/4 is perhaps most unique because of its very low accretion
rate. Our aperture-masking
observations of these systems (paper in preparation) can so far eliminate
stellar (mass ratio $q>0.1$) companions over a 20-160\,mas projected separation
range, but can say little about the possibility of closer
companions. Accurate ($\sigma \la1$\,km\,s$^{-1}$)  radial velocity
monitoring of these systems is 
required to determine if any of them harbor a close ($\la$4\,AU)
companion. Radial velocity monitoring is most applicable
to DM~Tau, where the cleared inner hole is expected to have only a
3\,AU radius \citep{Calvet05}.
\citet{Simon00} used resolved CO 2-1 observations of disks around
several T~Tauri stars to measure their masses directly. These
measurements would have resulted in the total mass of the stellar
primaries and any unresolved companions. The total masses of
$0.84\pm0.05$, $0.97\pm0.03$ and $0.55\pm0.03$ for GM~Aur, LkCa~15 and
DM~Tau respectively provide convincing evidence that these systems do
not harbor companions with mass ratios $q>0.3$.  

In summary, CoKu Tau/4
is a binary star system with a current projected separation of $\sim$7.8\,AU,
and a mass ratio near unity. The disk surrounding
CoKu Tau/4 is therefore a circumbinary disk, with the disk inner edge
set by dynamical truncation. The dynamical disk truncation radius of
$\ga13-16$\,AU is inconsistent with the disk radius derived from previous
spectral modeling, but it is likely that a different choice of
reddening law and correction, dust composition and geometry will
resolve this small discrepancy. We suggest 
%that most binary systems with separations
%similar to CoKu Tau/4 may be free of optically-thick disks, and
that systematic radial velocity monitoring and aperture masking
surveys are required to determine if other
so-called transition disks are in fact multiple systems. 

\acknowledgements

We gratefully acknowledge helpful discussions with Lynne Hillenbrand,
Klaus Pontoppidan, Geoffrey Blake, Gregory Herczeg and Colette
Salyk, and acknowledge the assistance of Thierry Forveille in pointing
out an oversight in an early version of the manuscript. M.I. would like to
acknowledge 
Michelson Fellowship support from the Michelson Science Center and the
NASA Navigator Program. A.L.K. is supported by a NASA/Origins grant to
L. Hillenbrand. 
The authors wish to recognize and acknowledge
the very significant cultural role and reverence that the summit of
Mauna Kea has always had within the indigenous Hawaiian community.  We
are most fortunate to have the opportunity to conduct observations
from this mountain. 

Facilities: \facility{Keck:II (NIRC2)}

\bibliography{../mireland}
\bibliographystyle{apj}

\end{document}